\documentclass[%
 reprint,
superscriptaddress,
preprintnumbers,
 amsmath,amssymb,
pre,
]{revtex4-2}

\usepackage{CJK}

\usepackage{amsmath}
\usepackage{physics}
\usepackage{braket}
\usepackage{graphicx}
\usepackage{dcolumn}
\usepackage{bm}

\usepackage{graphicx}
\usepackage{overpic}
\usepackage{afterpage}

\usepackage{xcolor} 
\usepackage[colorlinks=true]{hyperref} 
\newcommand{\ryd}{\mathbin{\bullet}} %
\newcommand{\gnd}{\mathbin{\circ}} 
\def\be{\begin{equation}}
\def\ee{\end{equation}}
\def\ba{\begin{eqnarray}}
\def\ea{\end{eqnarray}}
\def\bi{\begin{itemize}}
\def\ei{\end{itemize}}

\makeatletter
\def\frontmatter@title@above{\addvspace{2\p@}}
\def\frontmatter@title@below{\addvspace{1\p@}}
\def\frontmatter@authorformat{%
  \skip@\@flushglue
  \@flushglue\z@ plus.3\hsize\relax
  \centering
  \parskip2\p@\relax
  \@flushglue\skip@
}
\def\frontmatter@above@affiliation@script{%
  \skip@\@flushglue
  \@flushglue\z@ plus.3\hsize\relax
  \centering
  \@flushglue\skip@
  \addvspace{1\p@}%
}
\def\frontmatter@affiliationfont{%
  \footnotesize\itshape
  \setlength{\baselineskip}{9.2\p@}%
}
\def\frontmatter@preabstractspace{2\p@}
\def\frontmatter@postabstractspace{3\p@}
\def\frontmatter@finalspace{\addvspace{6\p@}}
\makeatother

\begin{document}


\title{Area-law entanglement in quantum chaotic systems  }


\begin{CJK}{UTF8}{gbsn}
\author{Chunyin Chen (陈椿尹)}
\thanks{These authors contributed equally to this work.}
\affiliation{International Center for Quantum Materials, School of Physics, Peking University, Beijing 100871, China}
\author{Sizhe Yan (闫思哲)}
\thanks{These authors contributed equally to this work.}
\affiliation{International Center for Quantum Materials, School of Physics, Peking University, Beijing 100871, China}
\author{Biao Wu (吴飙)}
\email{wubiao@pku.edu.cn}
\affiliation{International Center for Quantum Materials, School of Physics, Peking University, Beijing 100871, China}
\affiliation{Wilczek Quantum Center, Shanghai Institute for Advanced Studies, University of Science and Technology of China, Shanghai 201315, China}
\affiliation{Hefei National Laboratory, Hefei 230088, China}
\affiliation{Beijing Key Laboratory of Quantum Devices, Peking University, Beijing 100871, China}

\date{\today}

\begin{abstract}
Entanglement entropy is a fundamental diagnostic for quantum chaos, 
typically exhibiting volume-law scaling in highly excited eigenstates 
of chaotic many-body systems. In this work, we present a striking counterexample: 
a Floquet-driven quantum many-body system with a Rydberg-like blockade that, 
despite being fully chaotic as indicated by its Wigner-Dyson level statistics and local thermalization, 
exhibits a strict area-law entanglement entropy. 
Specifically, the entanglement entropy of every Floquet eigenstate is bounded by 
$\ln2$, independent of system size. We trace this anomaly to the specific Hilbert space structure imposed by the blockades, 
which restricts the Schmidt rank across a bipartition. Furthermore, 
we generalize this discovery by establishing a duality between constrained many-body Hamiltonians 
and single-particle quantum walks on median graphs, 
and we outline a general procedure for constructing systems with an entanglement entropy 
bounded by a predetermined constant. Our results demonstrate 
that entanglement entropy alone is an insufficient diagnostic of 
many-body quantum chaos and highlight the profound impact of Hilbert space geometry on quantum 
dynamics and thermalization.
\end{abstract}

\maketitle
\end{CJK}
\raggedbottom
\afterpage{\flushbottom}

\section{Introduction}

Entanglement entropy has emerged in recent years as a key diagnostic tool for tracking 
how quantum information spreads and scrambles throughout a quantum many-body 
system. In generic, chaotic quantum many-body systems, highly excited eigenstates 
are expected to exhibit volume-law entanglement entropy, meaning it scales linearly with the subsystem volume~\cite{vidmarEntanglementEntropyEigenstates2017d,bianchiVolumeLawEntanglementEntropy2022}.
Consequently, when a chaotic system is initialized in a simple product state, its entanglement 
entropy typically grows linearly before reaching this volume-law plateau.
This is the dynamical counterpart to Page's result for random states ~\cite{pageAverageEntropySubsystem1993}, 
where entanglement saturates near its maximal value.

Deviations from this universal scaling signal a breakdown of ergodicity. For instance, 
while some integrable systems~\cite{leblondEntanglementMatrixElements2019,jafarizadehBipartiteEntanglementEntropy2019,vidmarEntanglementEntropyEigenstates2017a,lydzbaEigenstateEntanglementEntropy2020,eisertColloquiumAreaLaws2010b,vidalEntanglementQuantumCritical2003,panCompositeSpinApproach2022,lydzbaEntanglementManybodyEigenstates2021} 
can exhibit volume-law entanglement, the coefficients are subsystem dependent and often lead to values 
below the maximum; others display distinctly sub-volume-law scaling. Similarly, in systems that weakly violate the eigenstate thermalization hypothesis (ETH) -- such as those hosting quantum many-body scars~\cite{chandranQuantumManyBodyScars2023a,hoPeriodicOrbitsEntanglement2019a,moudgalyaQuantumManyBodyScars2022g,turnerWeakErgodicityBreaking2018a,turnerQuantumScarredEigenstates2018e,serbynQuantumManybodyScars2021a,desaulesHypergridSubgraphsOrigin2022,bernienProbingManybodyDynamics2017f,windtSqueezingQuantumManyBody2022} or weak Hilbert space fragmentation~\cite{khemaniLocalizationHilbertSpace2020c,moudgalyaHilbertSpaceFragmentation2022b,salaErgodicityBreakingArising2020e,yangHilbertSpaceFragmentationStrict2020} -- the majority of states may follow a volume law, while a sparse, exponentially small subset of nonthermal states exhibits anomalously low, sub-volume-law entanglement.
In systems strongly violating the ETH, such as those exhibiting strong fragmentation\cite{morningstarKineticallyConstrainedFreezing2020,rakovszkyStatisticalLocalization2020,classenhowesUniversalFreezingTransitions2025}, the Hilbert space splits into many disconnected sectors, with both area-law and volume-law eigenstates coexisting in different sectors.

In this work, we present a Floquet-driven quantum many-body system with Rydberg-like interaction
where the entanglement entropy fails as a diagnostic of chaos. At a specific driving frequency, the system unambiguously exhibits quantum chaotic behavior: its level statistics follow the circular orthogonal ensemble, and local observables  in its Floquet eigenstates are indistinguishable from those in an infinite-temperature ensemble. Nevertheless, the bipartite entanglement entropy of all Floquet eigenstates is exactly or approximately $\ln 2$, independent of the system size.
This constitutes a direct violation of the expected volume-law scaling for a chaotic system.

The underlying mechanism for this anomaly is a Rydberg-like blockade that confines 
the system's dynamics to a special subspace, 
similar to the well-known PXP model\cite{bernienProbingManybodyDynamics2017f,turnerWeakErgodicityBreaking2018a,turnerQuantumScarredEigenstates2018e,serbynQuantumManybodyScars2021a,moudgalyaQuantumManyBodyScars2022g,sachdevMottInsulatorsStrong2002,fendleyCompetingDensityWave2004}. 
Within this subspace, whose dimension grows polynomially with the system size, the degenerate ground states 
possess a structure that strictly limits the Schmidt rank of the reduced density matrix, 
capping the entanglement entropy. This finding demonstrates that entanglement entropy 
alone is an insufficient metric for diagnosing many-body quantum chaos. 
Furthermore, we show that this system is a specific instance of a broader class of systems, 
which have either Rydberg or Rydberg-like blockades. 
These systems, which include the  PXP model, can be mapped to 
single-particle quantum walks on median graphs, and many share 
the property that their entanglement entropy is bounded by a constant, independent of system size.

\section{Anomalous Entanglement in a Chaotic Model}\label{main}
\subsection{The Constrained Rydberg-like Chain}\label{mainresult}
We study a Hamiltonian for a chain of atoms with Rydberg-like interaction:~\cite{bernienProbingManybodyDynamics2017f} 
\\
\begin{equation}\label{H0}
\begin{aligned}
\hat{H}= & \frac{\hbar}{2} \sum_{j=1}^{N}\left(\Omega(t) e^{i \varphi(t)}\ket{0}_j\bra{1}_j+\text{ h.c. }\right) \\-&\sum_{j=1}^{N} \hbar \Delta(t) \hat{n}_j 
+ V\sum_{j=1}^{N-1}  \hat{n}_j(1-\hat{n}_{j+1}).
\end{aligned}
\end{equation} 
where $\ket{0}_j$ represents the ground state and  $\ket{1}_j$  the Rydberg state at site $j$.  
$\hat{n}_j=\ket{1}_j\bra{1}_j$ is the number operator for the Rydberg state. 
$N$ is the total number of atoms. $\Omega(t)$, $\varphi(t)$, and $\Delta(t)$ are control parameters of 
a coherent laser beam. The last term with $V>0$ describes an interaction that is 
similar but  different from that in a conventional Rydberg-atom chain~\cite{HV}. 
Due to this interaction, for two nearest neighbors, the state where the left atom is in $\ket{1}$ 
and the right atom is in $\ket{0}$ is energetically unfavored. 
This kind of interaction may be engineered in experiments since it can be broken into 
two parts:  $V\hat{n}_j$ can be simply absorbed into the detuning term $\Delta(t)\hat{n}_j$; the rest
is an attractive interaction between two atoms in Rydberg state. The attractive interaction has been realized experimentally~\cite{dingErgodicityBreakingRydberg2024}, or it can be  simulated on Rydberg-array atoms platform with at most $O(N^2)$ atoms~\cite{nguyenQuantumOptimizationArbitrary2023a}.

\begin{figure}
    \centering
    \includegraphics[width=\linewidth]{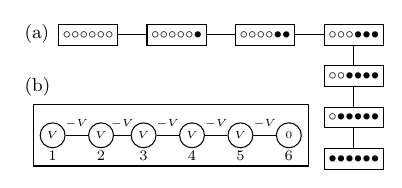}
    \caption{(a) The product states that span the constraint subspace for $N=6$. 
    $\gnd$ represents an atom in the ground state, while $\ryd$ represents an atom in the excited state. 
    Two product states are connected by a line when the Hamming distance between them is 1. 
   (b) The Rydberg-like interaction. The circle represents a Rydberg atom. The $V$  inside 
   the circle represents the energy of the atom in the Rydberg state, and the solid lines 
   represent the interactions between two atoms.
}
    \label{fig:solutionspace}
\end{figure}

We focus on the situation where $V\gg\abs{\hbar\Omega}$, $\abs{\hbar\Delta}$. 
In this case, similar to PXP model for Rydberg atoms, we can  project the Hamiltonian onto 
its subspace that is spanned by states favored by the interaction. 
As illustrated in Fig.~\ref{fig:solutionspace} with $N=6$, 
these states obey the constraint that for all $i<j$, $n_i\leq n_j$. 
Mathematically, such a subspace can be expressed as: 
\begin{equation}
    \mathcal{H} = 
    \operatorname{span}_\mathbb{C}\big\{ \ket{0^{N-k} 1^k } \;\big|\; k=0,1,\cdots, N \big\}\,,
\end{equation}
where $\ket{0^{N-k} 1^k }= \ket{0}^{\otimes(N-k)} \ket{1}^{\otimes k}$ is the so-called clock state~\cite{clock}.  
Within this subspace, the effective Hamiltonian closely resembles the PXP Hamiltonian. 
With Pauli operators defined as $\hat{Z}_j = 2\hat{n}_j - 1$, $\hat{X}_j = \ket{0}_j\bra{1}_j + \ket{1}_j\bra{0}_j$, 
$\hat{Y}_j = -i\ket{1}_j\bra{0}_j + i\ket{0}_j\bra{1}_j$, 
we can approximate the original Hamiltonian as: 

\begin{equation}\label{H_flq}
\begin{aligned}
\hat{H}_\text{flq}= 
& J(t) \sum_{j=2}^{N-1} P_{j-1}^{0} \left[ \cos{\varphi(t)}\hat{X}_j+\sin{\varphi(t)}\hat{Y}_j\right]P_{j+1}^{1}\\
+&J(t)\left[\cos{\varphi(t)}\hat{X}_1+\sin{\varphi(t)}\hat{Y}_1\right]P_{2}^{1}\\
+&J(t)P_{N-1}^{0} \left[\cos{\varphi(t)}\hat{X}_N+\sin{\varphi(t)}\hat{Y}_N\right]\\
+&A(t)\sum_{j=1}^N \hat{Z}_j,
\end{aligned}
\end{equation}
where $P_i^0$ and $P_i^1$ denote the projection operators onto the ground state $\ket{0}$ 
and the Rydberg state $\ket{1}$ at site $i$, respectively. 
The quantities $J(t)=\tfrac{\hbar}{2}\Omega(t)$, $A(t)=\tfrac{\hbar}{2}\Delta(t)$, and $\varphi(t)$ 
represent externally tunable, time-dependent control parameters.  

\subsection{Anomalous entanglement under Floquet driving}
\label{dynamics}

\begin{figure}[htbp]
  \centering
  \begin{minipage}[t]{0.22\textwidth}
    \centering
    \includegraphics[width=\linewidth]{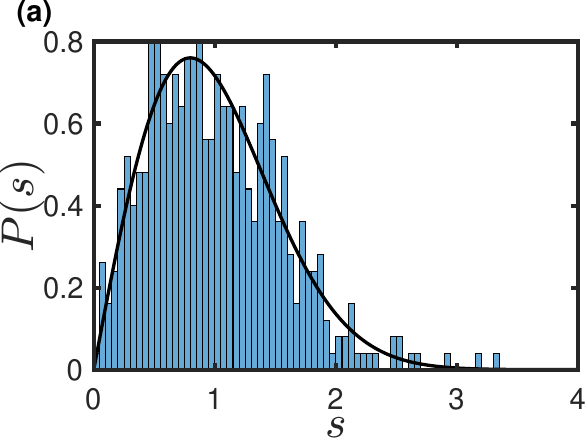}
    \label{fig:Ps}
  \end{minipage}
  \begin{minipage}[t]{0.23\textwidth}
    \centering
    \includegraphics[width=\linewidth]{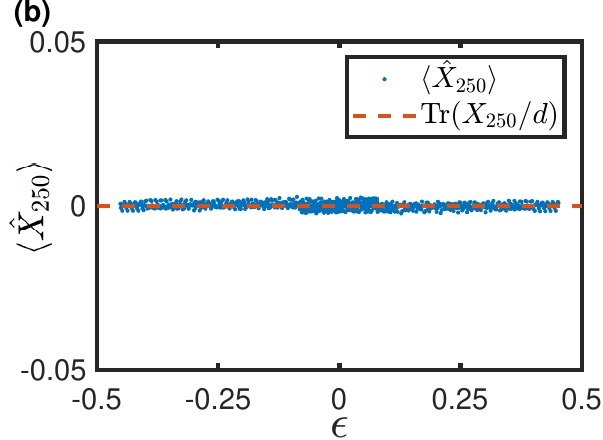}
    \label{fig:Xe}
  \end{minipage}
    \begin{minipage}[t]{0.23\textwidth}
    \centering
    \includegraphics[width=\linewidth]{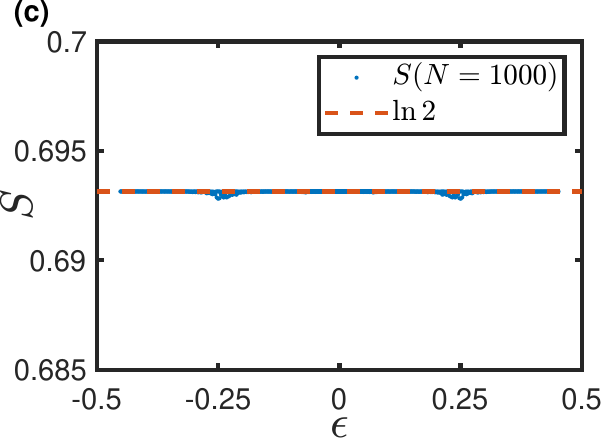}
    \label{fig:Se}
  \end{minipage}
  \begin{minipage}[t]{0.23\textwidth}
    \centering
    \includegraphics[width=\linewidth]{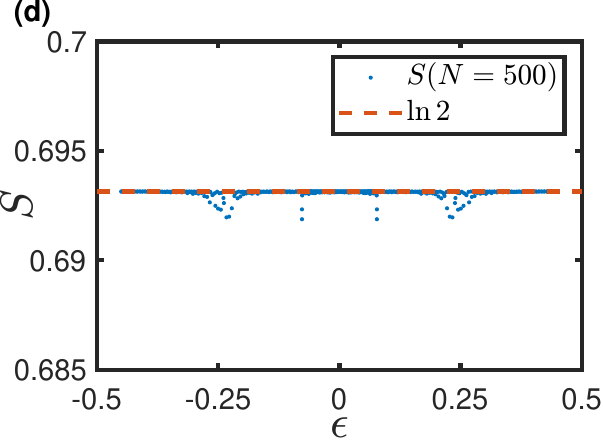}
    \label{fig:Se500}
  \end{minipage}
  
\caption{Indications of chaos  in our quantum many-body system with $\omega = 0.9071$.  
(a) Statistics of quasienergy level spacing for $N=1000$. $P(s)$ represents the probability distribution of the nearest-neighbor 
quasienergy level spacing $s$, fitted by a Wigner-Dyson distribution, with the solid line representing the standard COE level-spacing distribution. 
(b) Expectation value of the spin-flipping operator $\hat{X}_{250}$ at the 250th site in each Floquet eigenstate for $N=1000$, showing $\braket{\hat{X}_{250}} \approx 0$ for all eigenstates. 
(c) and (d) Entanglement entropy of the Floquet eigenstates for $N=1000$ (c) and $N=500$ (d), both clearly bounded by $\ln 2$. $\epsilon$ and $S$ refer to the quasienergy and von Neumann entropy, respectively. 
}\label{fig:main}

\end{figure}

Here we are interested in the potential chaotic behavior of our model, 
and we put it under Floquet driving, which is described as:
\begin{equation}
\begin{aligned}
   J(t) &= \sqrt{1+\cos^{2}(\omega t)},\\
   A(t) &= \cos(\omega t),\\
   \varphi(t) &= \omega t,
\end{aligned}
\end{equation}
where $\omega$ is the driving frequency. For the purpose of illustration, we choose $\omega = 0.9071$. The behavior of the system at different driving frequencies can be viewed in Appendix \ref{app:Scan}.
We have computed the quasienergy levels and related Floquet eigenstates. We find that its level-spacing statistics follows the 
Wigner-Dyson statistics, specifically the circular orthogonal ensemble (COE)~\cite{COE}, 
$P(s)=\frac{\pi}{2} s \exp \left(-\frac{\pi}{4} s^2\right)$, characterized by the level repulsion at $s\rightarrow 0$ 
and a Gaussian tail as 
shown in Fig.~\ref{fig:main}(a). It's a strong indication that our model with Floquet driving of frequency 
$\omega = 0.9071$ is a chaotic system\cite{floquetthermal}. This is further confirmed by our results on  
 local operator $\hat{X}_j$. As shown in Fig.~\ref{fig:main}(b),  our numerical computation finds that 
the expectation value  $\braket{\hat{X}_j}_n$ with respect to the $n$th Floquet eigenstate (labeled by the quasienergy $\epsilon$) 
is very close to $\mathrm{Tr}(\hat{X}_j)/d$, where $\mathrm{Tr}(\hat{X}_j)=0$ and $d=2$ is the dimension of the subspace of site $j$. 
This shows that every Floquet eigenstate is locally indistinguishable from an
infinite-temperature ensemble~\cite{infT}. This local thermal behavior, however, should not be confused with Haar randomness: the temporal ensemble
generated by a fixed chaotic Hamiltonian generally remains far from a unitary
design without additional perturbations~\cite{YangWu2026UnitaryDesigns}.

However, entanglement entropy fails to capture this strong chaotic behavior. We have computed the entanglement entropy (von Neumann entropy) for every Floquet eigenstate. As shown in Fig.~\ref{fig:main}(c,d), for two cases $N=500$ and $N=1000$, the entanglement entropy is either equal to $\ln 2$ or very close to $\ln 2$, completely independent of the system size. 

Generally, in chaotic many-body quantum systems, the bipartite entanglement entropy of typical high-energy eigenstates 
obeys the volume law. This means that 
the bipartite entanglement entropy of chaotic Floquet eigenstates should scale linearly with the subsystem size~\cite{volumeflq,volume1,volume2}.  This expectation fails in our model, where  
 the entanglement entropy remains essentially constant, saturating near $\ln{2}$, independent of the system size. 

The coexistence of chaotic Floquet behavior and system-size-independent entanglement entropy is atypical of previously studied quantum chaotic systems.
This behavior should also be distinguished from familiar nonergodic mechanisms.
In integrable systems, the entanglement entropy of excited states can obey volume-law scaling,
but the corresponding volume-law coefficient generally differs from the chaotic value and may depend on the subsystem fraction, namely the ratio between the subsystem size and the total system size~\cite{lydzbaEigenstateEntanglementEntropy2020, leblondEntanglementMatrixElements2019}.
In many-body localized (MBL) systems, highly excited eigenstates typically obey an area law, so in one spatial dimension the entanglement entropy across a bipartition remains $O(1)$.
By contrast, our model exhibits chaotic Floquet signatures while the entanglement entropy of its Floquet eigenstates remains bounded by $\ln 2$, independent of the system size.

We have examined other frequencies, and find that 
as $\omega$ is decreased from the high- to the low-frequency regime, the system exhibits a 
dynamical crossover from a prethermal plateau to full thermalization.  As they are not very relevant to this work, 
we will discuss them elsewhere. We only want to mention that when there is no Floquet driving, 
our model is integrable and  exhibits Bloch oscillations (for details, see Appendix \ref{app:IL}). 
 
That the entanglement entropy of our model is bounded by $\ln 2$ is due to the special structure 
of the states that span the subspace.  
Consider a chain of even length $N$. The Hilbert space of the half-chain subsystem is 
$\mathcal{H}_{L/R} = \operatorname{span}_\mathbb{C}\{ \ket{0^{\frac{N}{2}-k} 1^{k}}_{L/R} \mid k =0,1,\cdots,N/2 \} $, 
Any state in the  subspace can  be expressed in the coefficient-matrix form as
\begin{equation}\label{decomp1}
\ket{\psi}=\sum_{m,n=0}^{N/2} C_{mn}\ket{m}_L \otimes \ket{n}_R,
\end{equation}
where $\ket{m}_L=\ket{0^{\frac{N}{2}-m} 1^{m}}_{L}$ and $\ket{n}_R=\ket{0^{\frac{N}{2}-n} 1^{n}}_{R}$. 
The coefficients $C_{mn}$ form a matrix $C$.  
Due to the constraints enforced by the interaction, 
only elements with $m=0$ or $n=\tfrac{N}{2}$ are nonzero.  Thus, $C$ takes the form

\begin{equation}\label{cmatrix}
\begin{aligned}
C=\begin{pmatrix}
C_{00} & C_{01} & \cdots & C_{0,\frac{N}{2}-1} & C_{0,\frac{N}{2}}\\
0 & 0 & \cdots & 0 & C_{1,\frac{N}{2}}\\
\vdots & \vdots & \ddots & \vdots & \vdots\\ 
0 & 0 & \cdots & 0 & C_{\frac{N}{2},\frac{N}{2}} 
\end{pmatrix}. 
\end{aligned}
\end{equation}
The matrix elements correspond one-to-one with the subspaces in Fig. \ref{fig:solutionspace}(a). It is straightforward to see that the rank of $C$ is at most $2$, implying that the maximal entanglement entropy is $\ln 2$. 
This is why the entanglement entropy of our chaotic eigenstates is bounded above by $\ln{2}$.   
This observation underscores that entanglement entropy by itself is not a reliable diagnostic of integrability, nor is it sufficient to establish the presence (or absence) of many-body scars in the spectrum. Our analysis shows that 
a system's bipartite entanglement entropy is highly sensitive to the structure of its Hilbert space. 
When the Hilbert spaces of the two subsystems deviate markedly from a simple tensor-product structure, 
the volume-law scaling of entanglement entropy can fail even in a dynamically chaotic setting.

\section{General Construction of Models with Bounded Entanglement}
In this section, we present a general procedure to construct a quantum many-body system whose
entanglement entropy is bounded by a given value that is independent of the system size. 
The feasibility of such a construction is due to the fact that we are able to establish 
a duality between quantum many-body Hamiltonian and quantum walk on  median graph associated with 
a given 2-SAT problem. 

\subsection{Duality between quantum many-body Hamiltonian and quantum walk on  median graph}
\label{dualgraph}
The one-dimensional  system described in Eq.~(\ref{H0}) can be generalized to 
quantum  systems where atoms are arranged on a generic graph and interact 
with neighbors with Rydberg-like interaction. When the interaction is strong enough, these systems 
become confined to subspaces, where the dynamics can be regarded as single particle 
quantum walk on median graphs~\cite{chungDynamicLocationProblem1989a}. 
The well-known PXP model is a special case.  As we will see,  among these systems, 
there are many whose entanglement entropy in the constraint subspace is bounded by a finite value, 
independent of the system size.

The conventional Rydberg interaction between atoms has the following form: 
\be
U_{ij}=V \hat{n}_i\hat{n}_{j}\,,
\ee
with $V>0$. Among four possible states for these two atoms, $\{\ket{00},\ket{01},\ket{10},\ket{11}\}$, 
only $\ket{11}$ has positive interaction energy and all the other three have zero energy. In other words, if $V$ is big enough, 
the state $\ket{11}$ can be regarded as forbidden energetically. This is the well-known Rydberg blockade, which 
has been utilized to encode the problem of  independent set~\cite{HV}.  We regard this conventional Rydberg blockade as a special case of a broader class of Rydberg-like blockade terms, where a positive energy penalty is assigned to a selected two-site configuration. 

Independent set of a graph is a special type of 2-SAT problem~\cite{wuQuantumIndependentsetProblem2020a}, 
which is defined to find 
an assignment that satisfies a special type of constraints, namely
\begin{equation}
    C = \bigwedge_{(i,j)\in \text{edges}} (\neg x_i \lor \neg x_j) = 1\,,
\end{equation}
where $x_i$ is a binary variable and  $\neg x_i$ is its  negation. 
To encode all 2-SAT problems, we need to generalize the interaction as follows: 
 \begin{equation}
 	\begin{aligned}
 		 \neg x_i \lor \neg x_j &\Leftrightarrow   \hat{n}_i \hat{n}_j, \\
 		  \neg x_i \lor  x_j    &\Leftrightarrow  \hat{n}_i(1- \hat{n}_j), \\
 		  x_i \lor \neg x_j     &\Leftrightarrow   (1-\hat{n}_i) \hat{n}_j, \\
 		    x_i \lor  x_j       &\Leftrightarrow  (1- \hat{n}_i)(1-\hat{n}_j ).\\
  	\end{aligned}
 \end{equation}
 In the above, the left-hand sides are four types of clauses in 2-SAT problems. 
 The right-hand side are four corresponding Rydberg-like interactions. 
The correspondence between the clause and the interaction is rather obvious. For example, the second clause has three solutions $00, 01, 11$ and one non-solution $10$, which corresponds to three states $\ket{00}, \ket{01}, \ket{11}$ which are favored by the second interaction and one state $\ket{10}$ which is unfavored. This clause corresponds to the interaction term in Hamiltonian $\eqref{H0}$. Compared to the Rydberg blockade, which forbids the consecutive occurrence of 11, the prohibition of consecutive 10 significantly compresses the entire subspace.

With the above correspondence, we can construct a Hamiltonian with Rydberg-like interaction 
for any 2-SAT problem.  Consider an arbitrary 2-SAT problem 
with the following clauses
\begin{equation}
    C = \bigwedge_{(i,j)\in \mathcal{E}_{\mathcal C}} 
    (\tilde{x}_i \lor \tilde{x}_j) =1\,,
\end{equation}
where each $\tilde{x}_i$ denotes either $x_i$ or $\neg x_i$, and $\mathcal{E}_{\mathcal C}$ denotes the set of signed variable pairs appearing in the clauses, with the corresponding literals $\tilde{x}_i$ and $\tilde{x}_j$ specified for each pair. The Hamiltonian can be constructed as follows:
 \begin{equation}\label{H_2SAT}
\hat{H}_{\text{2-SAT}} = \frac{\Omega}{2} \sum_i \hat{X}_i - \Delta \sum_i \hat{n}_i + V \sum_{(i,j)\in \mathcal{E}_{\mathcal C}} \hat{\tilde{n}}_i \hat{\tilde{n}}_j
 \end{equation}
where $\tilde{n}_i$ denotes either $n_i$ or $1-n_i$.

When the interaction strength $V$ is big enough,  the quantum many-body system described by the 
Hamiltonian (\ref{H_2SAT}) can be regarded as living in the constraint Hilbert subspace spanned by
the solutions of the corresponding 2-SAT problem, 
\begin{align}
\label{2-SATsub}
     \mathcal{H}_\text{2-SAT} = \text{span}_\mathbb{C}\Big\{ \ket{n_1 n_2\cdots n_N} &\mid \tilde{n}_ i  \tilde{n}_j =0 , \\
    &\forall\  (i,j)\in \mathcal{E}_{\mathcal C} \Big\}. \nonumber
\end{align}
If  each solution is regarded as a vertex and two vertices are connected if and only if the Hamming distance between 
them is one, then these solutions form a median graph~\cite{chungDynamicLocationProblem1989a,zhaoQuantumHamiltonianAlgorithms2025a}. 
There is a one-to-one correspondence between 
a 2-SAT problem and a median graph: for a given 2-SAT problem, its solutions form a median graph; 
for a given median graph, we can always find a 2-SAT problem whose solutions form the given median graph.  
When the atoms are arranged in a line with real Rydberg interaction, this subspace is the Fibonacci cube. 

The effective Hamiltonian in this subspace is given by 
\be
\label{eq:Ha}
 \hat{H}_\text{eff} =-\Delta (t) D+\frac12\Omega (t) O\,,
\ee
where $D$ is a diagonal matrix with its $i$th element $D_i$ being the number of ones in the $i$th solution of the 2-SAT problem. 
$O$ is also a matrix, whose elements $O_{ij}$ are nonzero when the Hamming distance between the $i$th and $j$th solutions 
is one. The element of matrix $O$ are given by 
\begin{equation}
    O_{ij} =
    \begin{cases}
        1, & \text{if } \operatorname{Ham}(s_i, s_j) = 1,\ \ket{s_i}, \ket{s_j} \in \mathcal{H}_{\text{2-SAT}}, \\[4pt]
        0, & \text{otherwise.}
    \end{cases}
\end{equation}
where $\operatorname{Ham}(s_i,s_j)$ denotes the Hamming distance between product states $\ket{s_i}$ and $\ket{s_j}$.
 This Hamiltonian describes a quantum walk on the median graph of the 2-SAT solutions.

\subsection{Construction of quantum many-body Hamiltonians}
With the duality established above, we can design a procedure to construct  
quantum many-body Hamiltonians whose entanglement entropy is bounded by a value independent of the system size. 
The construction consists of three steps: 
\begin{enumerate}
\item Construct a coefficient matrix $C$ similar to Eq. \eqref{cmatrix},  which has a finite rank 
that is independent of the matrix size;
\item Build a graph according to the matrix $C$; if it is not a median graph, repeat step 1;
\item Find the 2-SAT problem according to the median graph obtained in the above step, and then 
construct a quantum many-body Hamiltonian.   
\end{enumerate}

We now  demonstrate this procedure with an example.  We consider a specific  matrix 
\begin{equation}
C_{3} =
\begin{pmatrix}
C_{00} & C_{01} & \cdots & C_{0, \frac{N}{2}-1} & C_{0, \frac{N}{2}} \\
C_{1,0} & C_{1,1} & \cdots & C_{1, \frac{N}{2}-1} & C_{1, \frac{N}{2}} \\
\vdots & \vdots & \ddots & \vdots & \vdots \\
0 & 0 & \cdots & 0 & C_{\frac{N}{2}-1, \frac{N}{2}} \\
0 & 0 & \cdots & 0 & C_{\frac{N}{2}, \frac{N}{2}}
\end{pmatrix}
\end{equation}
which are formed by the coefficients  of the quantum states
\begin{equation}
\ket{\psi}=\sum_{m,n=0}^{N/2} C_{mn}\ket{m}_L \otimes \ket{n}_R\,.
\end{equation}
The rank of the matrix $C_3$ is at most 3, independent of the matrix size $N/2$. \\

\begin{figure}[h]
	\centering
	\includegraphics[width=0.55\linewidth]{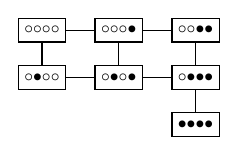}
	\caption{A median graph with $N=4$. The entanglement entropy of the 
	corresponding quantum system is bounded by $\ln 3$.}
	\label{Fig:line_ln3}
\end{figure}

According to the matrix $C_3$, we set up a graph. 
One example with $N=4$ is illustrated in Fig.~\ref{Fig:line_ln3}. 
It is straightforward to verify the so-constructed graph is a median graph. 
Since there always exists a 2-SAT problem for a given median graph, 
we are able to find the corresponding 2-SAT problem and construct the following  Hamiltonian:
\begin{align}
H  =&\frac{\Omega}{2} \sum_i \hat{X}_i - \Delta \sum_i \hat{n}_i + V
\hat{n}_{\frac{N}{2} - 1} (1 - \hat{n}_{\frac{N}{2} + 1})
\nonumber\\
&+V \sum_{i=1, i \ne N/2}^{N-1} \hat{n}_i (1 - \hat{n}_{i+1})\,.
\end{align}
Similar to the Hamiltonian in Eq.~(\ref{H0}),  this Hamiltonian also describes a chain of atoms. 
The difference is that there is no interaction between atoms at sites $\frac{N}{2}$ and  $\frac{N}{2}+1$ 
while there is interaction between atoms at sites $\frac{N}{2}-1$ and  $\frac{N}{2}+1$. 
When $V$ is large enough, the dynamics of this system is confined in a subspace where the largest 
entanglement entropy is $\ln 3$.  One may start directly with a median graph and construct a many-body Hamiltonian. 
However, this does not guarantee that the entanglement entropy of the system satisfies the area law.

\section{Conclusion}
We have presented a quantum many-body system, whose entanglement entropy is always bounded by $\ln 2$ 
even when it becomes chaotic under Floquet driving. This defies a long-held belief that
the highly excited eigenstates of a quantum many-body chaotic system should obey the volume law. 
Furthermore, we have outlined a procedure to construct more similar quantum many-body systems, 
whose entanglement entropy obeys the area law. Following this procedure, it may be possible to construct
two-dimensional quantum many-body systems whose entanglement entropy is bounded by a fixed value. 

The thermodynamic properties of such quantum many-body systems are expected to
be quite different from the conventional systems. Consider a scenario where 
such a system is thermalized by the heat bath with a temperature much smaller than $V$. 
In this case, the system is still confined to the subspace so that its 
thermodynamic entropy should be proportional to $\ln N$. This means that 
its thermodynamic entropy is much bigger than the entanglement entropy, offering 
a strong argument that the thermodynamic entropy is in general not equivalent  to the entanglement entropy\cite{Hu2019PRE}.  In addition, this result implies that 
such a thermodynamic entropy  is not extensive as it becomes  $\ln (2N)$ not $2\ln N$ 
when we double the system size. We will leave this for future discussion.

\begin{acknowledgments}
C.C. would like to thank Xianjue Zhao and Zhigang Hu for helpful discussions. 
This work was supported by the National Natural Science Foundation of China (92365202, 12475011, and 11921005), 
the National Key R\&D Program of China (2024YFA1409002), and 
the Shanghai Municipal Science and Technology Project (25LZ2601100).
\end{acknowledgments}

\appendix

\section{Our model without Floquet driving}\label{app:IL}
In the absence of Floquet driving, the system (\ref{H0}) becomes integrable and all its eigenstates can be found analytically. 
Without driving, the Hamiltonian \eqref{H_flq} reduces to the following form with $A(t)=0$
\be \label{H_1dchain}
H_{\text{hop}} = \sum_{j=0}^{N-1} \ket{j}\bra{j+1} + \text{h.c.}
\ee
where $\ket{j}=\ket{0^{N-j}1^j}$. It describes a single-particle hopping model 
on a one-dimensional chain with open boundary conditions. The above Hamiltonian is also 
a special form of the Hamiltonian in Eq.(\ref{eq:Ha}). In the following, we take $N$ to be  even.

This Hamiltonian has exact solutions. The eigenstates are given by
\begin{eqnarray}
	\ket{\phi_m} = -\sqrt{\frac{2}{N+2}} \sum_{j=0}^{N} \sin\!\left(\frac{(j+1)m\pi}{N+2}\right) \ket{j},  
\end{eqnarray}
with eigenvalues $E_m = 2 \cos\!\left(\frac{m\pi}{N+2}\right)$, where $m = 1, 2, \cdots, N+1$. 

We compute the bipartite entanglement by tracing out the right $N/2$ Rydberg atoms. 
For the eigenstates $\ket{\phi_m}$, let $k = \frac{m\pi}{N+2}$. 
The elements of the density matrix are given by
\be
\begin{aligned}
& \left(\rho_{1/2}\right)_{00} = \frac{2}{N+2}\sum_{j=0}^{N/2} \sin^2\!\left[k(j+1)\right], \\
& \left(\rho_{1/2}\right)_{ij} = \frac{2}{N+2}\sin\!\left[k\!\left(\frac{N}{2} + i + 1\right)\!\right] 
  \sin\!\left[k\!\left(\frac{N}{2} + j + 1\right)\!\right]\,,
\end{aligned}
\ee
where $i$ and $j$ are not equal to $0$ simultaneously. 

It can be readily checked that $\rho_{1/2}$ is a rank-$2$ matrix. 
When $m$ is even, we have $(\rho_{1/2})_{00} = \frac{1}{2}$ and $(\rho_{1/2})_{0i} = (\rho_{1/2})_{i0} = 0$. 
Therefore, the entanglement entropy is exactly $\ln 2$. 
When $m$ is odd, $\lim_{N \to \infty} (\rho_{1/2})_{00} = \frac{1}{2}$. 
Therefore, the entanglement entropy for all eigenstates approaches $\ln 2$ in the large-$N$ limit. 

We put back the last term in the Hamiltonian \eqref{H_flq}
\be \label{H_1dchain2}
H_{\text{hop}} = \sum_{j=0}^{N-1} \ket{j}\bra{j+1} + \text{h.c.}+A(t)\sum_{j=1}^N \hat{Z}_j\,. 
\ee
When $A(t)$ is a nonzero constant~\cite{Yu_2021}, the system exhibits Bloch oscillations~\cite{hartmannDynamicsBlochOscillations2004b}.

\section{Our system at different driving frequencies}\label{app:Scan}
In both the low- and high-frequency limits, the system is expected to exhibit approximately integrable dynamics. Let \(\phi=\omega t\). For \(\omega\to0\), the Hamiltonian \(H(\phi)\) varies slowly in time. When the adiabatic condition is satisfied, the Floquet operator becomes approximately diagonal in the adiabatic basis:
\begin{equation}
    U(T)\ket{n(0)}
\simeq
\exp\left[
-\frac{i}{\omega}\int_0^{2\pi}E_n(\phi)d\phi
+i\gamma_n
\right]\ket{n(0)} .
\end{equation}
The low-frequency dynamics is therefore governed mainly by adiabatic phase accumulation, with the adiabatic quantum number approximately conserved, leading to effectively integrable behavior.
In the high-frequency limit \(\omega\to\infty\), the Floquet dynamics is governed by an effective Hamiltonian,
\begin{equation}
    U(T)=\exp[-iH_{\mathrm{eff}}T],
\;\;
H_{\mathrm{eff}}
=
\frac{1}{T}\int_0^T H(t)dt.
\end{equation}
The cycle-averaged Hamiltonian has an integrable structure and the high-frequency Floquet operator is also approximately integrable. In particular, when the averaged term is weak, the quasienergy phases may collapse, causing the spectrum to deviate from chaotic circular-ensemble statistics.

Consequently, chaotic behavior is expected mainly at intermediate frequencies. In this regime, nonadiabatic transitions strongly mix Floquet eigenstates and destroy approximate conserved quantities. The quasienergy statistics then depart from the integrable behavior and approach the circular random-matrix ensemble determined by the system symmetry, such as COE.

\begin{figure}[htbp]
  \centering
  \begin{minipage}[t]{0.35\textwidth}
    \centering
    \includegraphics[width=\linewidth]{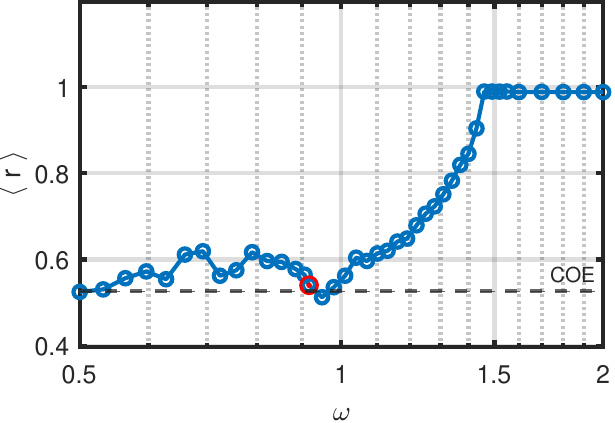}
  \end{minipage}
\caption{The average adjacent-gap ratio \(\langle r\rangle\) over \(\omega\in[0.5,2]\). The red dots represent the selected representative frequency, and the dashed lines represent the values of \(\langle r\rangle\) corresponding to COE. It can be seen that phase collapse occurs at high frequencies.
}\label{Fig:scan}
\end{figure}

We compute the average adjacent-gap ratio \(\langle r\rangle\) over \(\omega\in[0.5,2]\) to select a representative intermediate-frequency parameter, as shown in Fig.~\ref{Fig:scan}. This window does not access the strict low-frequency adiabatic limit: for very small \(\omega\), the period \(T=2\pi/\omega\) becomes long, and constructing the Floquet operator accurately requires many time slices. Thus, the low-frequency side of the plot should only be viewed as a finite-frequency trend.

Within this scan window, \(\omega=0.9071\) lies in the intermediate-frequency regime and corresponds to a clear deviation of \(\langle r\rangle\) from regular static-spectrum behavior. It is therefore used as a representative point for intermediate-frequency dynamics. Since this frequency is not too low, the period remains moderate and the numerical evolution is relatively stable.

\providecommand{\noopsort}[1]{}\providecommand{\singleletter}[1]{#1}%

\end{document}